\documentclass[11pt]{article}
\usepackage{wrapfig}
\usepackage{epsfig}
\usepackage{subfigure}

\begin{document}

\title{Simulation of Pickup Signal in a
Resistive Plate Chamber}

\author{N. Majumdar, S. Mukhopadhyay, S. Bhattacharya}
\date{INO Section, Saha Institute of Nuclear Physics\\
 1/AF Bidhannagar, Kolkata - 700064\\
\small{nayana.majumdar@saha.ac.in, supratik.mukhopadhyay@saha.ac.in, 
sudeb.bhattacharya@saha.ac.in}}
\maketitle

\begin{abstract}
The signal induced by an electron on a pickup strip has been calculated
in a RPC used in INO calorimeter \cite{INOColl} following
Ramo's theorem \cite{Ramo39}. An accurate estimation of weighting field 
has been obtained using a nearly exact Boundary Element Method (neBEM)
solver \cite{Majumdar06} while the electron velocity from the real
field values.
The calculation has shown a growth and subsequent fall of pickup signal 
with duration about $90$ps.
\end{abstract}

\section{Introduction}
The resistive plate chamber (RPC) has become an integral part of modern high
energy experinments owing to its simple design and construction, good time
resolution, high efficiency and low cost production. Several 
problems like inherent rate limitation and much debated
space charge effect have induced various studies on detailed
simulation of RPC signals considering different physics and 
electronics aspects.

The induction of signal in the electrodes of a chamber can be treated
by Ramo's theorem following which
the electrostatic and weighting fields turn out to be 
two fundamental quantities. 
In general, instead of carrying out
detailed computation for realistic RPC geometries of
severl layers of dielectrics, the weighting potential and field
of a RPC is determined for a simpler geometry of three layers 
for which 2D analytic solutions can be derived \cite{Heubrandtner02}.
A thorough study on the 3D field distribution is required 
for realistic configuration of a RPC in order to achieve
a true estimation of the induced signal.
The field computation is usually carried out 
using commercial Finite Element Method (FEM) packages, although, it is
known to yield poor accuracy despite of rigorous computation.
A precise computation of 3D field distribution has been carried out 
using a nearly exact Boundary Element Method (neBEM).
In this formulation, many of the drawbacks of the
standard BEM like singularities and near-field inaccuracies could be
removed due to use of analytic solutions of potential and field.
A simple calculation of induced signal due to
the motion of an electron in the chamber volume has been carried out
with precise computation of 3D weighting
and actual field for actual geometry of an RPC to be used in INO calorimeter.

\section{Signal Induction Process}
According to Ramo's theorem, the current $i(t)$ induced 
due to a charge cluster $e_0N(t)$
moving with a drift velocity 
$\vec{v_D}(t) = \dot{\vec {x}}(t)$ can be evaluated as follows:
\begin{equation}
i(t) = e_0N(t)\vec{v_D}(t).\vec{E_w}(\vec{x}(t))
\end{equation}
where $\vec{E_w}(\vec{x}(t))$ is the
weighting field at $\vec{x}(t)$ associated with the electrode under study. 
It may be mentioned here that
the weighting field can be obtained when
readout electrode is set to $1$V keeping all other electrodes grounded. 

\section{Results and discussions}
\subsection{Comparison with analytical solutions}
\begin{figure}[htb]
\centering
\subfigure{\label{fig:WeightFFAnal3L}\includegraphics[width=0.45\textwidth]
{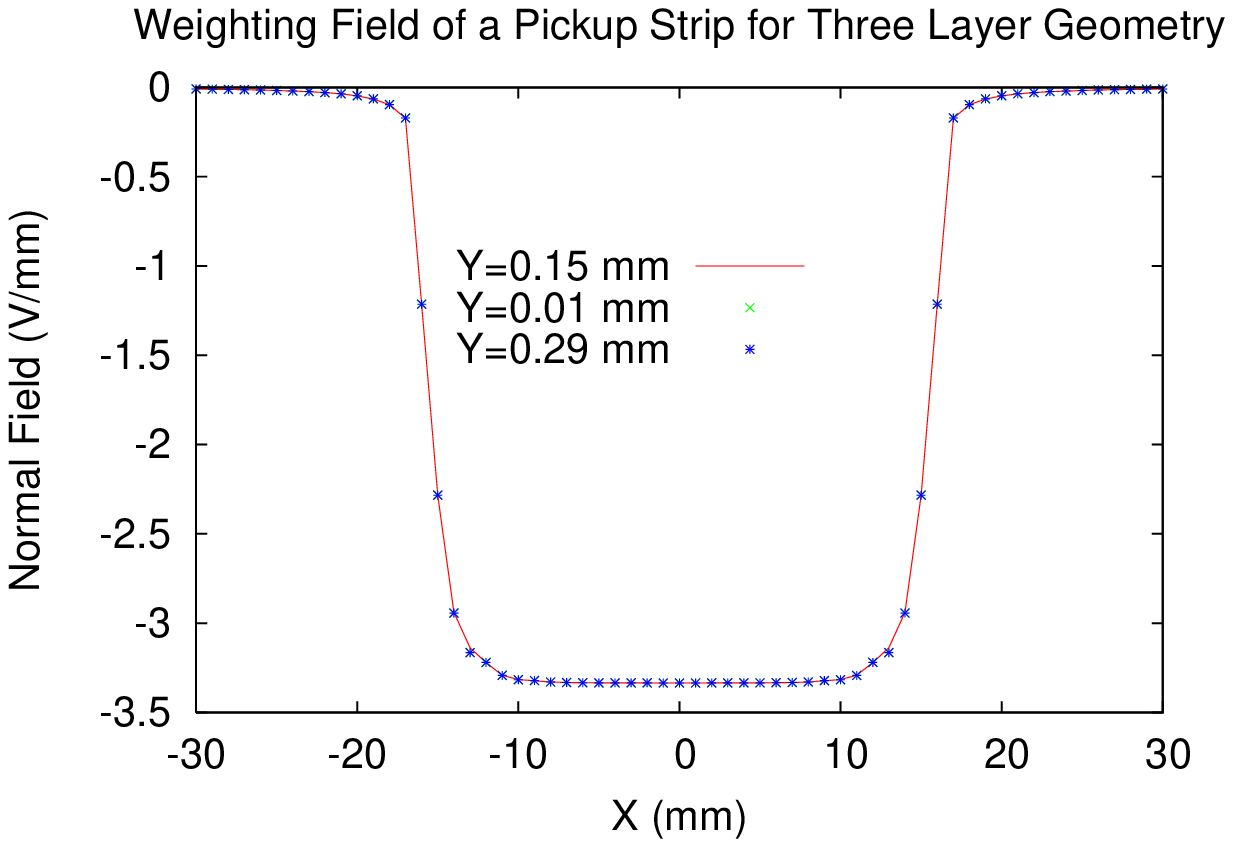}}
\subfigure{\label{fig:WeightFFanal}\includegraphics[width=0.45\textwidth]
{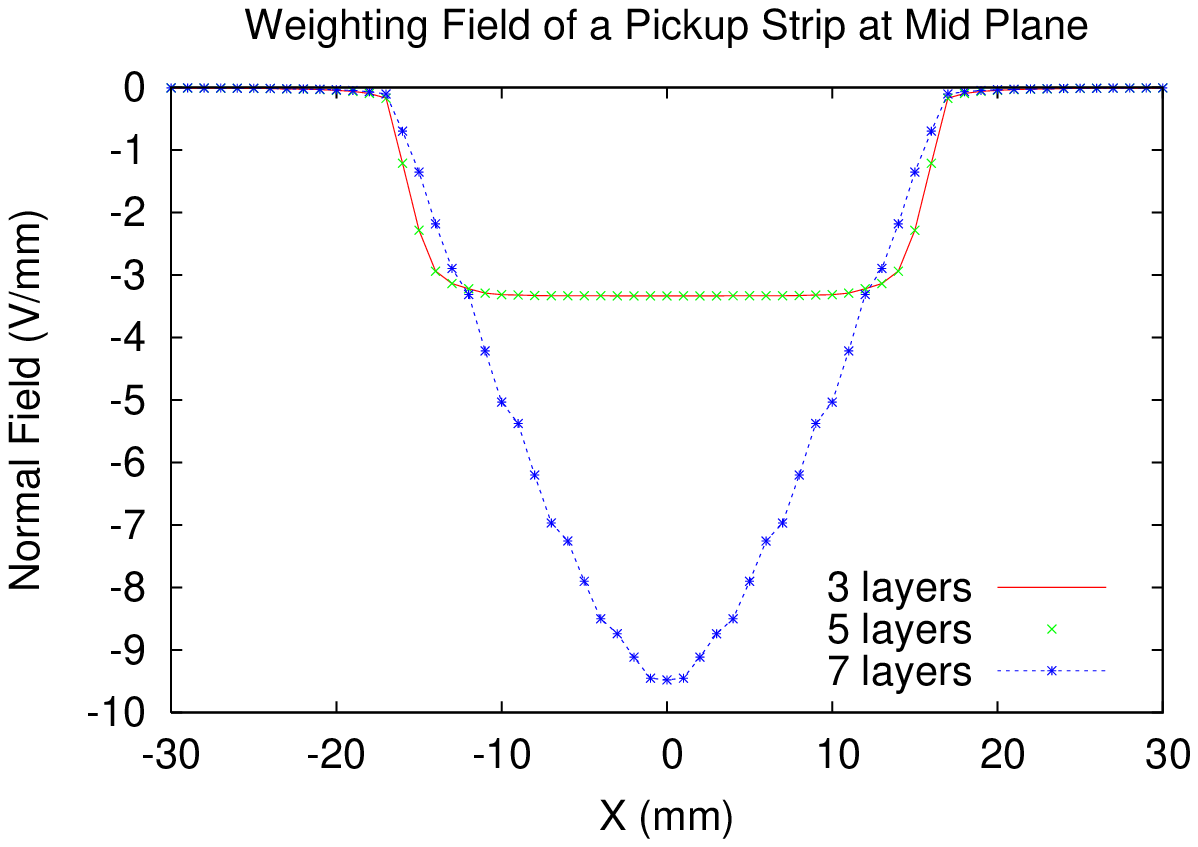}}
\caption{Analytic weighting field distribution of
(a) three layer and (b) multiple layer geometries}
\label{fig:WeightAnal}
\end{figure}
The results obtained for three layer INO RPC following analytic solutions
and neBEM calculations have been illustrated in fig.
\ref{fig:WeightAnal}(a) for three different positions along the axis of
the geometry. 
The analytic expressions have been found to be valid up to two more 
resistive layers. However, inclusion
of further layers has brought in a significant change (see 
fig.\ref{fig:WeightAnal}(b)) indicating that the naive use of analytic 
expressions may lead to
non-negligible errors.

\subsection{Comparison with other methods}
The solver has been validated by computing the electrical properties of
a configuration containing two layers of dielectrics with a thin layer
of graphite on each outer side to provide high voltage. 
\begin{table}
\begin{center}
\caption{Comparison for R = 10}
\begin{tabular}{|l|c|c|c|}
\hline \textbf{Location} & \textbf{FEM} & \textbf{DBEM} & 
\textbf{NEBEM}\\ \hline
18.0,3.0  & 0.1723103 & 0.17302 & 0.1740844 \\
4.0,9.0   & 0.2809692 & 0.27448 & 0.2807477 \\
25.0,16.0 & 0.6000305 & 0.59607 & 0.5991884 \\
5.0,17.0  & 0.679071  & 0.67492 & 0.6785017 \\ \hline
\end{tabular}
\label{table:t1}
\end{center}
\end{table}
The potentials
at different locations
within the dielectrics following standard formulations \cite{Chyuan04} 
and neBEM have been tabulated in table \ref{table:t1}, R being the ratio 
of dielectric constants of upper to lower plate.

\subsection{Calculation for INO RPC}
\begin{figure}[htb]
\centering
\subfigure{\label{fig:3DWeightFFINO7L}
\includegraphics[width=0.45\textwidth]{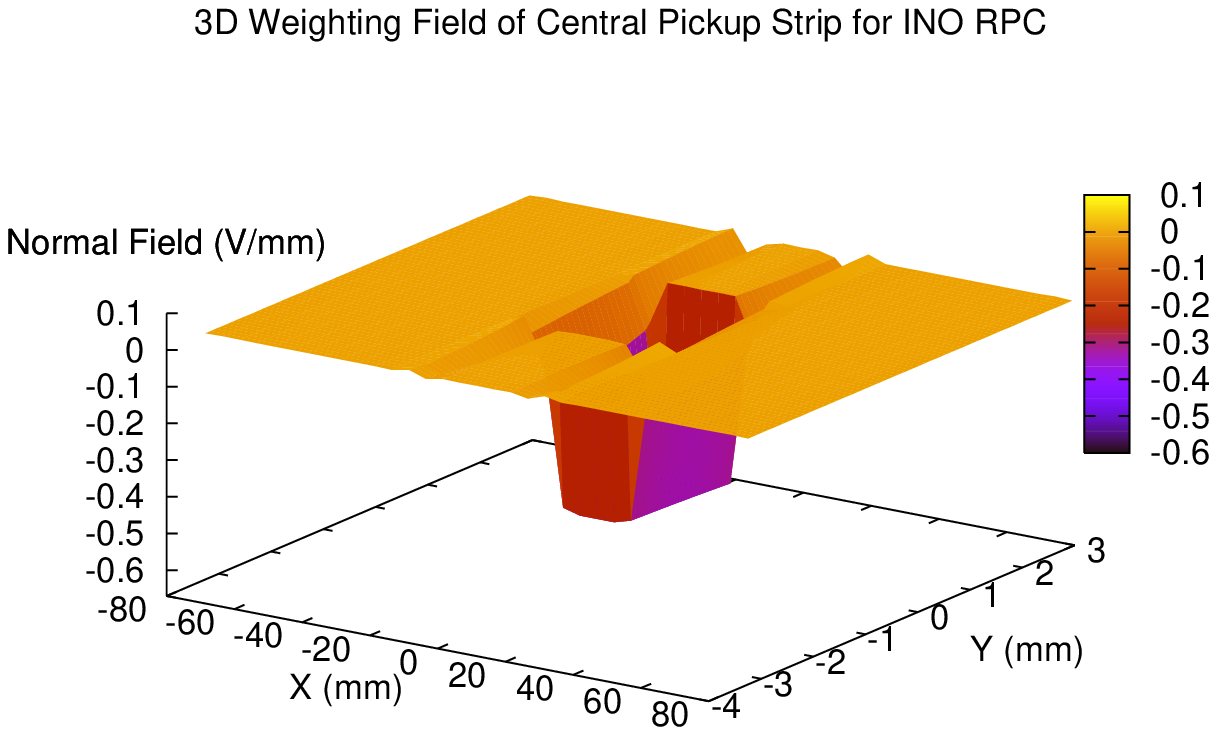}}
\subfigure{\label{fig:3DActualFFINO7L}
\includegraphics[width=0.45\textwidth]{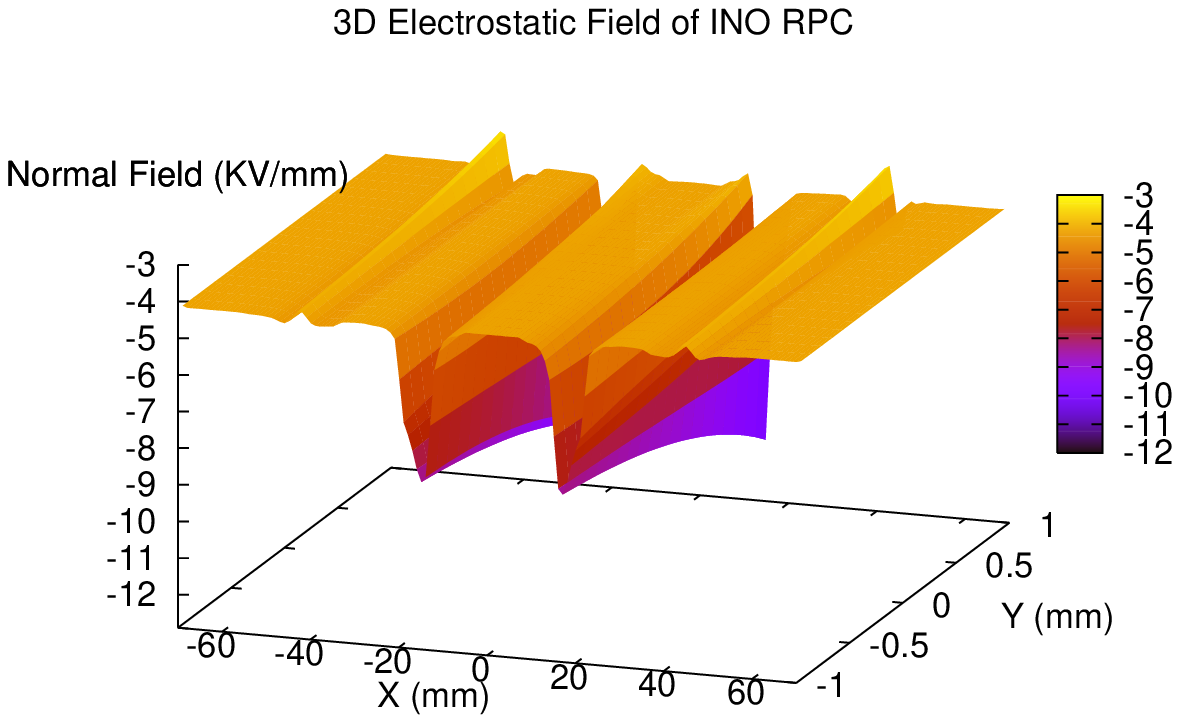}}
\caption{Three dimensional distribution of (a) weighting field and 
(b) actual field of the RPC in INO Calorimeter}
\label{fig:3DField}
\end{figure}

A glass RPC with glass and gas thickness of $2$mm has been considered.
A thin graphite coating of thickness $50\mu$m has been provided with a
high voltage of $8.5$KV. The readout strips are separated from the graphite
layers by a PET film, $100\mu$m thick.
Five readout strips in X-direction have been considered with thickness
$16\mu$m, width $3$cm and pitch $1$mm while the Y-readout has been 
considered to be an uniform plate of same thickness. The length of the 
chamber in Z-direction has been made $50$cm.
However the inclusion of PET film above the graphite layer
has been found to
generate some numerical instability and excluded afterwards.
The dielectric constants of glass, graphite and gas
layers have taken to be $7.75$ (float glass), $12$ and 
$1.000513$ (Argon) respectively. The results of
the weighting and real field distributions have been shown 
in fig.\ref{fig:3DField}(a) and (b). 

\subsection{Calculation of induced signal}
\begin{wrapfigure}[14]{r}{0.5\textwidth}
\begin{center}
\includegraphics[width=0.45\textwidth]{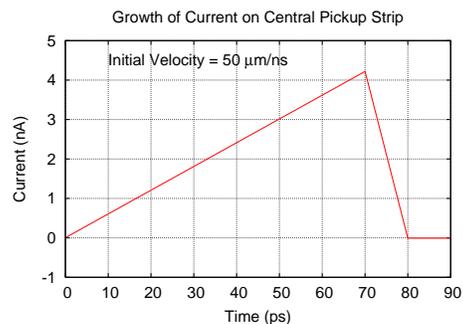}
\caption{\label{fig:InduceCurrent} Induced current due to the motion
of an electron in INO RPC}
\end{center}
\end{wrapfigure}

The induced current has been calculated at intervals of $10$ps during
the passage of the electron in the gas gap. The electron has been assumed
to pass perpendicularly across the detector plane at its center with
a velocity $50\mu$m/ns \cite{Riegler04} starting from an initial 
position $10\mu$m above the lower glass layer.
The weighting field has been evaluated at each position of
the electron due to the central readout strip while the velocity has
been determined from the real field using neBEM. The result is plotted in 
fig.\ref{fig:InduceCurrent} which has shown the growth of the signal while
electron has moved in the gas gap and a subsequent fall
as the electron has left the gap. 
A rough estimate of the intrinsic timing resolution \cite{Riegler03} with
the values of Townsend and attachment coefficients following the 
detector simulation package GARFIELD
\cite{Garfield} for the electric field as estimated by the neBEM has
turned out to be about $1\over4$ns.

\section{Conclusion}
Using the neBEM solver, the three dimensional weighting potential and field
can be calculated precisely even for a very detailed geometry of a RPC.
This allows us to simulate the induced current on any electrode of a RPC 
due to the passage of an ionizing particle. The drift velocity of the
particle can be 
calculated using GARFIELD.
However, the inclusion of several very thin layer of dielectrics
has been found to generate numerical instabilities for which the solver would
have to be optimized.

\end{document}